\documentclass[final]{aa}
\usepackage{graphicx,epsfig,fancyhdr,rotating,amsmath,natbib}
\usepackage{txfonts,url}
\def\etal {{ et al. }}

\newcommand{\degree}{\ensuremath{^\circ}}
\usepackage{txfonts}

\begin{document}
\title{On the statistical detection of propagating waves in polar coronal holes}
\author{G. R. Gupta\inst{1,2} E. O'Shea\inst{3}\and D. Banerjee\inst{1} \and M. Popescu\inst{3} \and J.G. Doyle\inst{3}}
\institute{Indian
Institute of Astrophysics, Bangalore 560034, India \and Joint Astronomy Programme, Indian Institute of Science, Bangalore 560012 \and Armagh Observatory, College Hill, Armagh BT61 9DG,
N.Ireland }
\date{\today}

\abstract 
{Waves are important to the study of dynamical processes in coronal holes and the acceleration of the fast solar wind. A spectral time series was taken with the SUMER spectrometer on-board SoHO on 20 October 1996. The observations were obtained in the N~{\sc iv} 765 \AA\ transition region line and the Ne~{\sc viii} 770 \AA\ line of the low corona.}
{We detect the presence of waves and study their characteristic
  properties in terms of their propagation speeds and
  direction. Previous statistical studies, undertaken with data from
  the CDS spectrometer, report the presence of waves in these
  regions. We extend this analysis using SUMER observations.}
{Using Fourier techniques, we measured the phase delays between
 intensity oscillations, as well as between velocity oscillations, in
 our two lines over the full range of available frequencies.
 From this, we were able to measure the travel time of the propagating 
oscillations, hence the propagation speeds of the waves that produce the
 oscillations.}
{We detect the long period oscillations in polar coronal holes on the disc. For network bright locations 
within coronal holes, our
 results indicate the presence of compressional waves with a dominant 
 period  of $\approx$25 min.  However, we also find power at many
 other different frequencies, so we are able to study oscillations
 over a full range of frequencies. We find evidence of propagating
 waves with a fixed time delay in the coronal hole. We find, moreover,
 that there is a difference in the nature of the wave propagation in the bright (`network'), as opposed to the dark (`internetwork') regions, with the
 latter sometimes showing evidence of downwardly propagating waves
 that are not seen in the former. From a measurement of propagation
 speeds, we find that all measured waves are subsonic in nature. }
{Waves with different characteristics are found to be present at
  different locations in the observed coronal hole. The measured
  propagation speeds are subsonic, indicating that the majority of
  them are slow magneto-acoustic in nature. These waves, measured in the lower
  atmosphere, could accelerate farther at higher altitudes and may be 
important for the acceleration of the fast solar wind.}

\keywords{Sun: corona - Sun: oscillations - Sun: UV radiation - Sun: transition region - Waves}

\titlerunning{On the statistical detection of propagating waves in polar coronal holes}
\authorrunning{Gupta \etal}

\maketitle

\section{Introduction}
Coronal holes are regions of cool and low density plasma that are
`dark' at coronal temperatures \citep{Munro72}.
During the years of solar minimum, coronal holes are confined
to the Sun's polar regions, while at solar maximum they can also be found
at lower latitudes, usually associated with remnant active regions, as
so-called `equatorial' coronal holes \citep{Timothy75}. The predominantly unipolar
magnetic field from coronal hole regions is thought to give rise
to the fast solar wind \citep{Krieger73}.

The prominent feature of the coronal holes in the chromosphere and transition region is the magnetic network, which is believed to be the upward extension of the supergranular boundaries above the photosphere \citep{Reeves1976}. The network is identified as bright lanes in radiance images, the darker areas within the network cells are termed as internetwork. It is suggested that part of the network fans out in the corona into a funnel shape. Recently, it
has been further suggested that the fast solar wind originates from coronal hole
funnels \citep{Tu05}. Different
studies have found evidence for outflows (blue shifts) of typically 10 
km s$^{-1}$ in both polar \citep{Wilhelm00, Popescu04} and equatorial
\citep{Xia04} coronal holes at transition region
temperatures.  

A number of studies have measured oscillations in coronal holes in the polar off-limb regions of
the Sun \citep{Ofman97, Ofman00, Banerjee01, Popescu05}. All of
these studies point to the presence of compressional waves, thought to
be slow magneto-acoustic waves as found 
by \citet{DeForest98}, \citet{O'Shea06, O'Shea07}.

In this work, we look for additional evidence of waves in the disc part of polar coronal
hole regions by carrying out a statistical
study of oscillations and their phase properties, both in the intensity
and velocity domain. The observations were carried out with the SUMER
spectrometer ( Solar Ultraviolet Measurements of Emitted
Radiation) on-board the Solar and Heliospheric
Observatory (SoHO) spacecraft \citep{Wilhelm95}. We will focus on the behaviour of oscillations detected in the bright network regions and the darker internetwork regions. 

\section{Observations and data analysis}
\begin{figure}[htbp]
\centering
%{\includegraphics[bb=240 750 545 90, width=5cm,angle=90, clip=]{lcurve_198_159.eps}}
{\includegraphics[bb=548 140 300 709, angle=90, width=9cm, clip=]{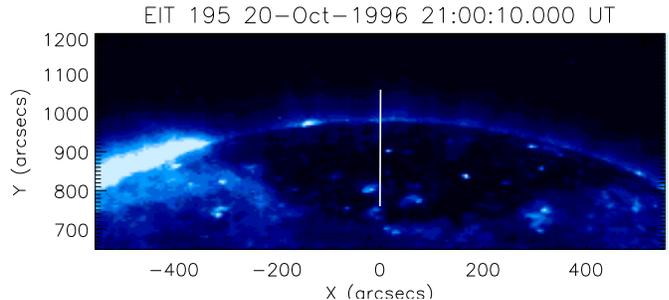}}
\vspace*{3.5cm}
 \caption{Solar polar coronal holes imaged by EIT/SoHO in the
 $195$ \AA\ band. The vertical white lines show the fixed position of the SUMER slit.}
 \label{fig:1a}
\end{figure}

The data selected for this study were obtained as a time series on 20 October 1996, starting at 19:57 UT, 
in a Northern polar coronal hole (hereafter pCH) as observed by SUMER
with the \textbf{$1\arcsec\times300\arcsec $} slit using an  exposure time of 30
seconds. The two lines used in
this study are the transition region N~{\sc iv} 765 \AA\ line at a
temperature of log T=5.0 K (T = 100 000 K) and the Ne~{\sc viii} 770 \AA\ coronal line at a temperature of log T=5.8 K (T = 630 000 K).
During the observation, the SUMER slit
was kept fixed at the north pole ($X=0\arcsec, Y=909.75\arcsec$) (see Fig.~\ref{fig:1a}) with part of the slit off-limb and part in the disc part of the coronal hole (sit-and-stare observations). Fig.~\ref{fig:1a} shows the initial position of the SUMER slit overplotted on an EIT (the Extreme
ultraviolet Imaging Telescope on the SoHO spacecraft) image taken in the $195$ \AA\ filter. The pCH can be
clearly identified in the image as very dark regions of reduced coronal emission at the poles. The slit width ($1\arcsec$) gives the
spatial resolution along the x-direction, whereas the resolution along the
slit in the y-direction (north-south; positive towards north) is given
by the pixel size of the detector, and is also approximately
$1\arcsec$. For sit-and-stare observations one needs to consider the
effect of the solar rotation on the detectable minimum frequency. 
We used the program, rot$_{-}$xy.pro within solarsoft~\footnote{\url{http://www.astro.washington.edu/deutsch-bin/getpro/library32.html?ROT_XY()}}, 
which incorporates diff$_{-}$rot.pro
\footnote{\url{ http://www.astro.washington.edu/deutsch-bin/getpro/library32.html?DIFF_ROT()}},
to calculate the rotation rate. The diff$_{-}$rot.pro procedure uses standard equations for
calculating solar rotation,
while rot$_{-}$xy.pro converts the results of this from degrees/day to
arcsec/day (second). At the lowest pixel location,
Y$\approx$760", the rotation rate is $\approx 3.7\arcsec$/hour.  Thus the highest
frequency that can possibly be affected by rotation is $ \approx 1.03
$ mHz \citep{Doyle98}. Whereas at high latitudes (Y=950" and higher) the solar rotation is very slow
($\leq \approx0.255\arcsec$/hour) and thus the spreading
of the frequencies is negligible ($\leq \approx 0.071$ mHz). In general, most frequencies are measured from
pixels above Y$\approx$760" and, thus, the effects of rotation will be 
at values much less than $ \approx 1.03$ mHz.

We applied the standard procedures for
correcting and calibrating the SUMER raw data, namely decompression, reversal as well as flat field, dead-time, local-gain, and geometrical corrections
\footnote{\url{ http://www.ias.u-psud.fr/website/modules/content_sol/index.php?id=1068}}. 
The method used to deduce the line parameters (radiance, central position of the spectral line and width) is described as follows. We calculated the central
position for every pixel, by integrating the line radiance
across a certain spectral window and determining subsequently
the location of the 50 \% level with sub-pixel accuracy. This procedure
is frequently used to obtain SUMER Dopplergrams (see
details in \citet{Dammasch99}) and dramatically reduces the
computing time for a large number of data. The results deduced
by this method are statistically consistent with those obtained
by using the standard Gaussian fitting programme \citep{Xia03}.
In addition, a line-position correction was applied, to remove
spurious spectral line shifts caused by thermal deformations of
the instrument, and to eliminate residual errors (systematic
variation along the slit) after the geometric correction, using the
standard software \citep{Dammasch99, Xia03}. The uncertainty in velocity from this calculation is $\pm 1.2$ Km/s \citep{Dammasch99}.
The line of sight (LOS) velocity values presented in this
paper are measured relative to the limb velocity, i.e., the velocity
at the limb is assumed to be effectively zero. We binned over two time frames and two spatial pixels to increase the signal to noise, which yields an effective 
cadence of $60$ s. The Nyquist frequency, therefore, is
$\approx 8.33 $ mHz \citep{Jenkins68}. The total observation time 
is $ 240$  min $=  1.44 \times 10^{4} $ s and, hence, the frequency
resolution is $\approx 6.94 \times 10^{-2} $ mHz. In the
following Fourier analysis, phase delays will be
measured at all the frequencies up to the Nyquist frequency, at steps
dictated by the frequency resolution.

%%%%%%%%%%%%%%%%%%%% figure 1
%----------------------------------------------
\begin{figure}[htbp]
\centering
\vspace*{-1.0cm}
 \includegraphics[angle=90, width=9cm]{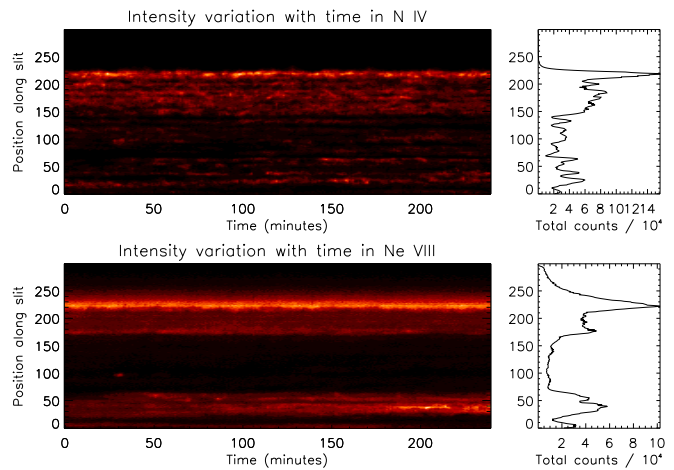}
\caption{Top (bottom) left panel shows the variation of the N~{\sc iv} 765 \AA\ (Ne~{\sc viii} 770 \AA\ ) intensity along the slit (in pixel units) with time. Right panel shows the
  variation of the total summed counts over time along the slit.}
\label{fig:1c}
\end{figure}
%%%%%%%%%%%%%%%%%%%%%%%%%%%%%%%%%%%%%%%%%%%%%%%%%%%%%%figure 2 end

The Quiet Sun (QS) chromosphere and transition region show a network
 pattern all over the disc, with intensity
enhancements in the network boundaries. 
Within coronal holes such patterns also exist
\citep{Tu05}. Presumably, the magnetic field is predominantly
concentrated on such network boundaries and, within coronal holes, the 
footpoints of coronal funnels emanate from these network regions. Recent studies from SOT/HINODE also indicate that these funnels originate from
small magnetic patches with very strong kG magnetic fields \citep{Tsuneta08}. \citet{Tsuneta08} further claims that all the open field lines forming the polar coronal hole 
essentially originate from such scattered small but intense magnetic patches, and the fast solar winds emanate from these vertical flux tubes seen in the photosphere. We assume that our 
intensity enhanced regions could correspond to these magnetic patches. Thus, to identify such locations and distinguish between
network and internetwork pixels, the subsets of data to be used 
in our study were selected on the basis of bright (`network')
and dark (`internetwork') locations.  The left panels of
Fig.~\ref{fig:1c} show the variation of the N~{\sc iv} (top panel) and the Ne~{\sc viii} (lower panel) intensity along the slit with time. The right panels show the variation of the total
summed counts over time at each pixel position along the slit. The intensity enhanced locations marks the location of network boundary. For
choosing the bright (`network') locations, we have used the
arbitrary criterion that pixels having an intensity higher than $1.35$
times the average intensity ($\sim$ 42000 counts) are network pixels (excluding
those very close to the limb and off-limb pixels, beyond $\approx$210
pixels). This corresponds to the intensity enhancements in the right top panel of
Fig.~\ref{fig:1c} and their nearest neighbours. The pixels numbered
$23-26$ and $\approx 149-200$  are
identified as bright pixels and other pixels as dark (internetwork) for the low temperature N~{\sc iv} line (note that in this particular case the
network boundary spans a large spatial domain along the slit, maybe it is a bigger patch as seen by recent HINODE studies).
The network pixels obtained from N~{\sc iv} line are assumed to be the same in the
higher temperature Ne~{\sc viii} line. 
%------------------------------------------------
\begin{figure*}[htbp]
\centering
\hspace*{-0.7cm}\includegraphics[angle=90, width=11cm, clip=]{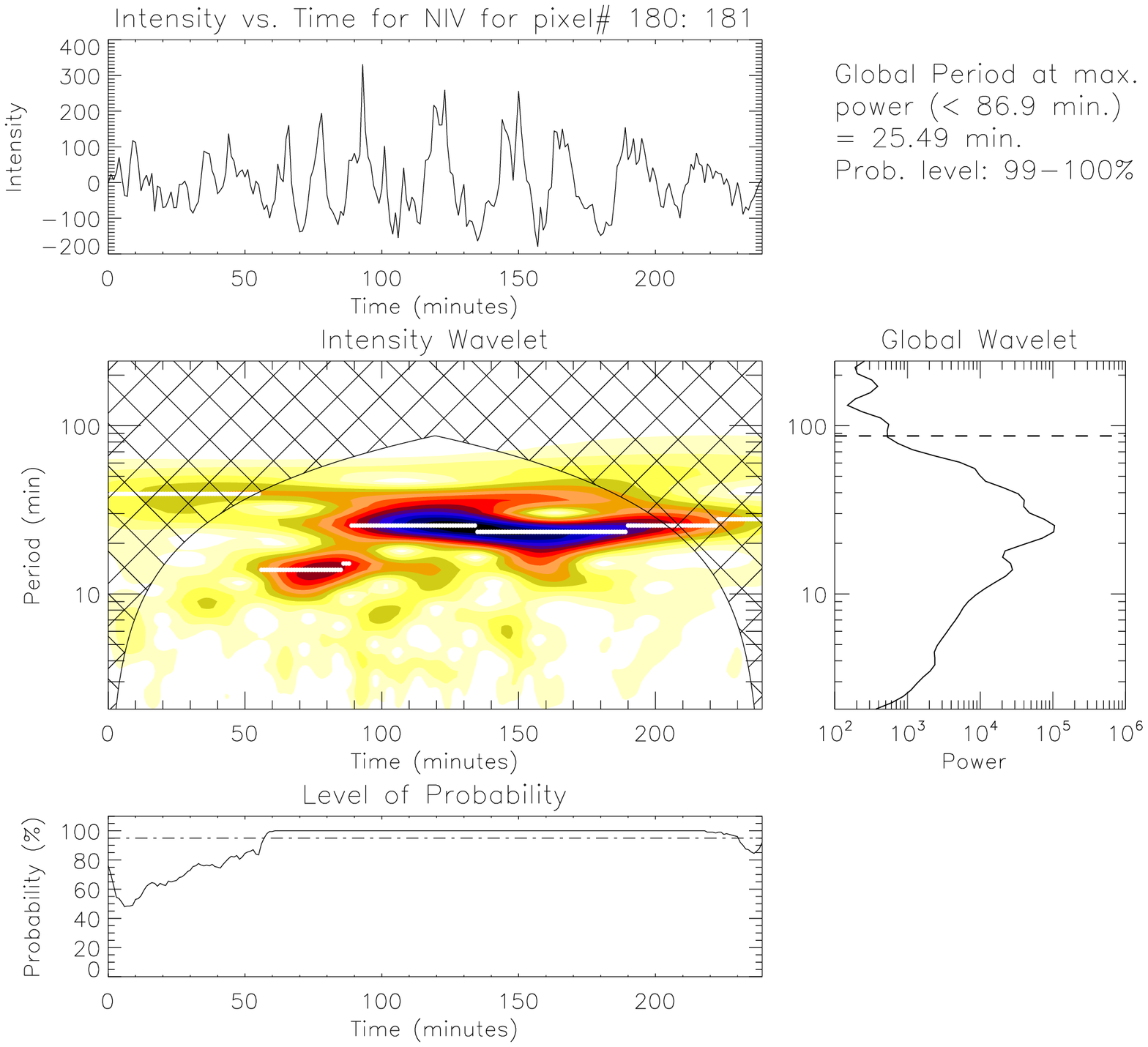}\hspace*{-1.5cm}{\includegraphics[angle=90, width=11cm, clip=]{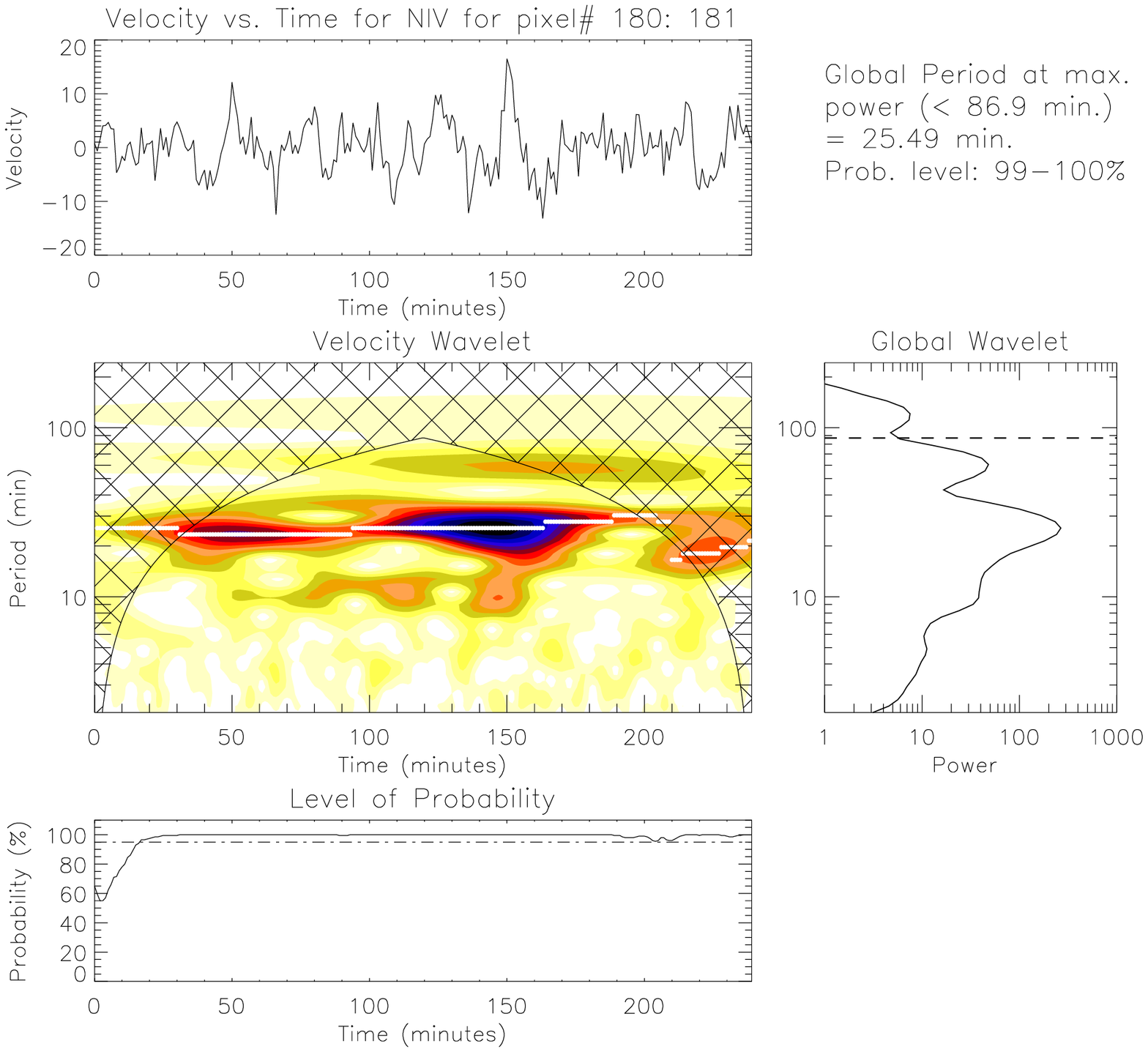}}
 \caption{The wavelet result for a bright location (pixel no. $=
 180-181$ which corresponds to Y$\approx940\arcsec$) in the  N~{\sc iv} 765 \AA\ intensity and velocity. In each set the
 top panels show the relative (background trend removed) radiant
 flux/velocity as marked, the central panels show the colour inverted wavelet power
 spectrum, the bottom panels show the variation of the probability
 estimate associated with the maximum power in the wavelet power
 spectrum (marked with white lines), and the right middle panels show the
 global (averaged over time) wavelet power spectrum. Above the global
 wavelet the period, measured from the maximum power from the global wavelet,
 together with probability estimate, is printed.}
 \label{fig:2a}
\end{figure*}
%-------------------------------------------------------------
%----------------------------------------------------------
\begin{figure*}[htbp]

\centering
\hspace*{-0.7cm}\includegraphics[angle=90, width=11.cm, clip=]{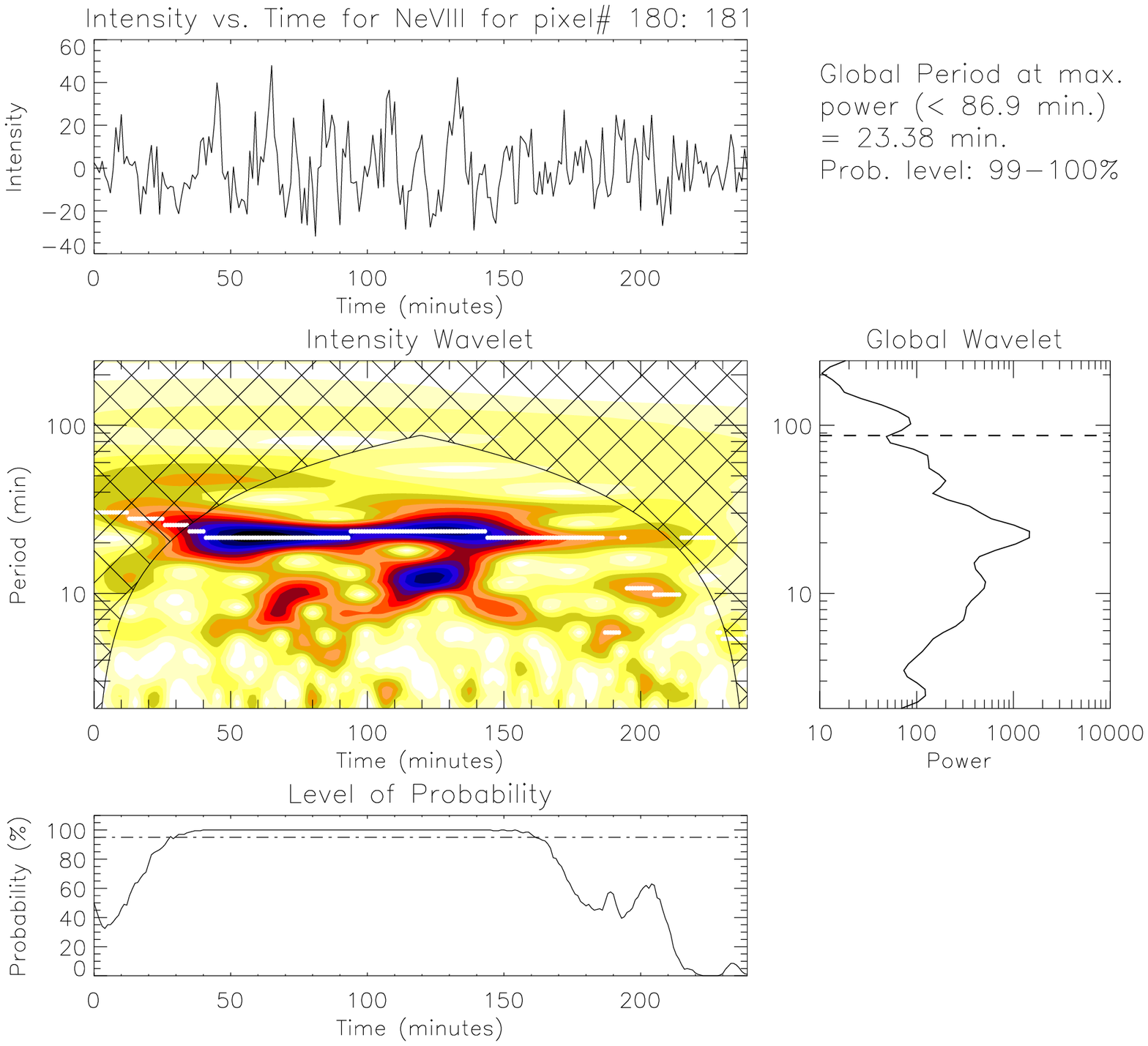}\hspace*{-1.5cm}{\includegraphics[angle=90, width=11.cm, clip=]{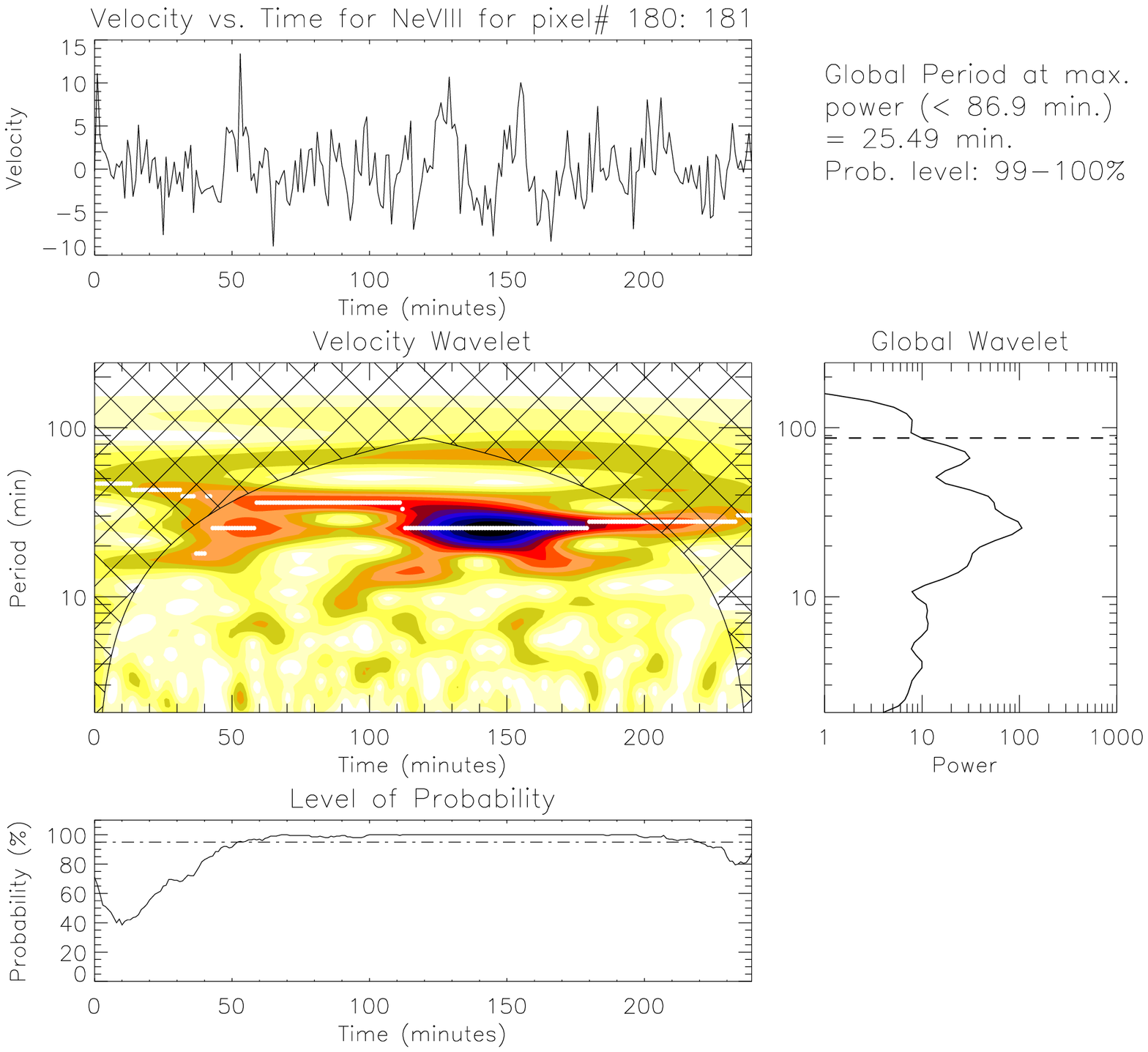}}
\caption{The wavelet analysis results for a bright location (pixel
 no. $= 180-181$ which corresponds to Y$\approx940\arcsec$) in the intensity (left side) and in velocity (right side) data for the
 Ne~{\sc viii} 770 \AA\ line. The descriptions about the different panels are the same as in Fig.~\ref{fig:2a}.}
 \label{fig:2c}
\end{figure*}
%-------------------------------------------------------
%\vspace*{7cm}
\section{Results and Discussions}
In Figs.~\ref{fig:2a} \&~\ref{fig:2c}, we show a representative example
of the type of the oscillations we measured in a bright region of the pCH at the pixel location 180-181 which corresponds to the solar-Y location of $\approx940\arcsec$. At this location the rotation rate is $\approx 0.87\arcsec$/hour.  Thus the highest
frequency that can possibly be affected by rotation is $ \approx 0.24$ mHz.
Fig.~\ref{fig:2a} shows the variation of the intensity (hereafter we will use the term intensity for trend subtracted radiant flux) and relative velocity (trend subtracted), as a function of
time. Details on the wavelet analysis, which provides information on
the temporal variation of a signal, are described in
\citet{Torrence98}. For the convolution with the time series in the
wavelet transform, we chose the Morlet function, as defined in 
\citet{Torrence98}. The oscillations shown in the upper panel had their
background trend removed by subtracting from the original time series a
$30-$point running average (i.e., a $\approx$30 min interval). In the 
wavelet spectrum, the
 cross-hatched regions are locations where estimates of oscillation period become unreliable. This is the
 so-called cone-of-influence (COI), see \citet{Torrence98}. As a result of the COI, the maximum measurable period is shown by a dotted line in the global spectrum plots. Above the global wavelet spectrum of  Fig.~\ref{fig:2a}  is shown the period, measured at the location of the
maximum in the global wavelet spectrum, together with an estimate of the
probability that this oscillation is not due to noise. 
 The
probability estimate was calculated using the randomisation method with $200$ permutations as outlined in detail in \citet{O'Shea01}.
A randomisation test is based on the assumption
that, if there is no periodic signal in the time series data,
then the measured values (intensity, velocity, etc.) are independent
of their observation times. For example, the
intensities \textbf{$I_{1}$, $I_{2}$,... $I_{n}$}, observed at times \textbf{$t_{1}$, $t_{2}$,... $t_{n}$}, are
just as likely to have occurred in any other order \textbf{$I_{r(1)}$,
$I_{r(2)}$,... $I_{r(n)}$}, where n is the total number of observations
and r(1), r(2),... r(n) is a random permutation of the subscripts
1, 2,... n. Only those
oscillations with a probability $>$95\% are considered to be
significant. Below the wavelet power spectrum, in the lower panels, we
show the variation of the probability estimate, calculated using the
randomisation technique, associated with the maximum power at each
time in the wavelet power spectrum. The location of the maximum power
is indicated by the overplotted white lines. We should also point out that there are 240 points in the intensity
time series, with a separation of 60s, which result in a range of
(non-linear) period scales, or values, given by the wavelet. These
scales (or period values) from the wavelet are at fixed values of period
and so small variations of period between one oscillation period and
another can be lost. In our case, in Figs.~\ref{fig:2a} \& ~\ref{fig:2c}, we see periods at
a reduced range of values of 25.49 and 23.38 min. It is most probable
that the oscillations have a range of values between 23-26 min. However,
their precise values must remain unknown due to the relative crudity of
the temporal resolution given by the wavelet scales. Thus we can see oscillations
of $\approx$ 25 min period in both the intensity and the
velocity. So, we may say that these oscillations are likely to be
present due to the propagation characteristics of a compressional wave.
These oscillations can be seen in both lines (N~{\sc iv} and
Ne~{\sc viii}), cf. Figs.~\ref{fig:2a} \& \ref{fig:2c}. We note that
the fact we see a similar periodicity in the intensity and velocity wavelets is a sign 
of `correlation' at the strongest period of oscillation. As the
oscillations are seen in both the lower and the higher temperature
lines, it is likely that the waves producing these oscillations are
propagating between the different temperature regions where these
lines form. 

To investigate whether this is actually the case, we measure the phase
delays in intensity and also in the LOS relative velocity
between both lines for each of the measurable pixels along the
slit. For all separate frequencies  the `correlation' is determined by the
squared coherency spectral estimate, which is a function of frequency, with
values between 0 and 1 that indicates how well a time series x
corresponds to a time series y at each frequency. The phases are calculated from cross power spectral estimates,
following the techniques outlined in the appendix of \citet{Doyle99}. Only phases (corresponding to
each spatial pixel) where the squared coherencies are greater than a
significance of 95.4\% (2 $\sigma$) are used in the analysis
(see \citet{O'Shea06} and references therein). The errors in the phase is calculated based on the equation (A23) of \citet{Doyle99}.
In this work we follow the treatment used by \citet{Athay79} and
\citet{O'Shea06}, in which the calculated phase delays are plotted
over the full $-180$ to
$+180$ (360$\degree$) range and as a function of the measured
oscillation frequency. Since the expected phase delay or difference is
given by the equation:
\begin{equation}
 \Delta\phi = 2\pi\textit{f}T
\end{equation}
where $\textit{f}$ is the frequency and T the time delay in seconds,
the phase difference will vary linearly with \textit{f}, and will
change by $360$ over frequency intervals of $\Delta \textit{f} =
1/T$. This will give rise to parallel lines in $\Delta\phi$
vs. $\textit{f}$ plots at fixed frequency intervals ($\Delta
\textit{f} =1/T$), corresponding to a fixed time delay T. 
\begin{figure*}[htbp]
 \centering
%\hspace*{10 cm}
\includegraphics[angle=90, width=8.7cm]{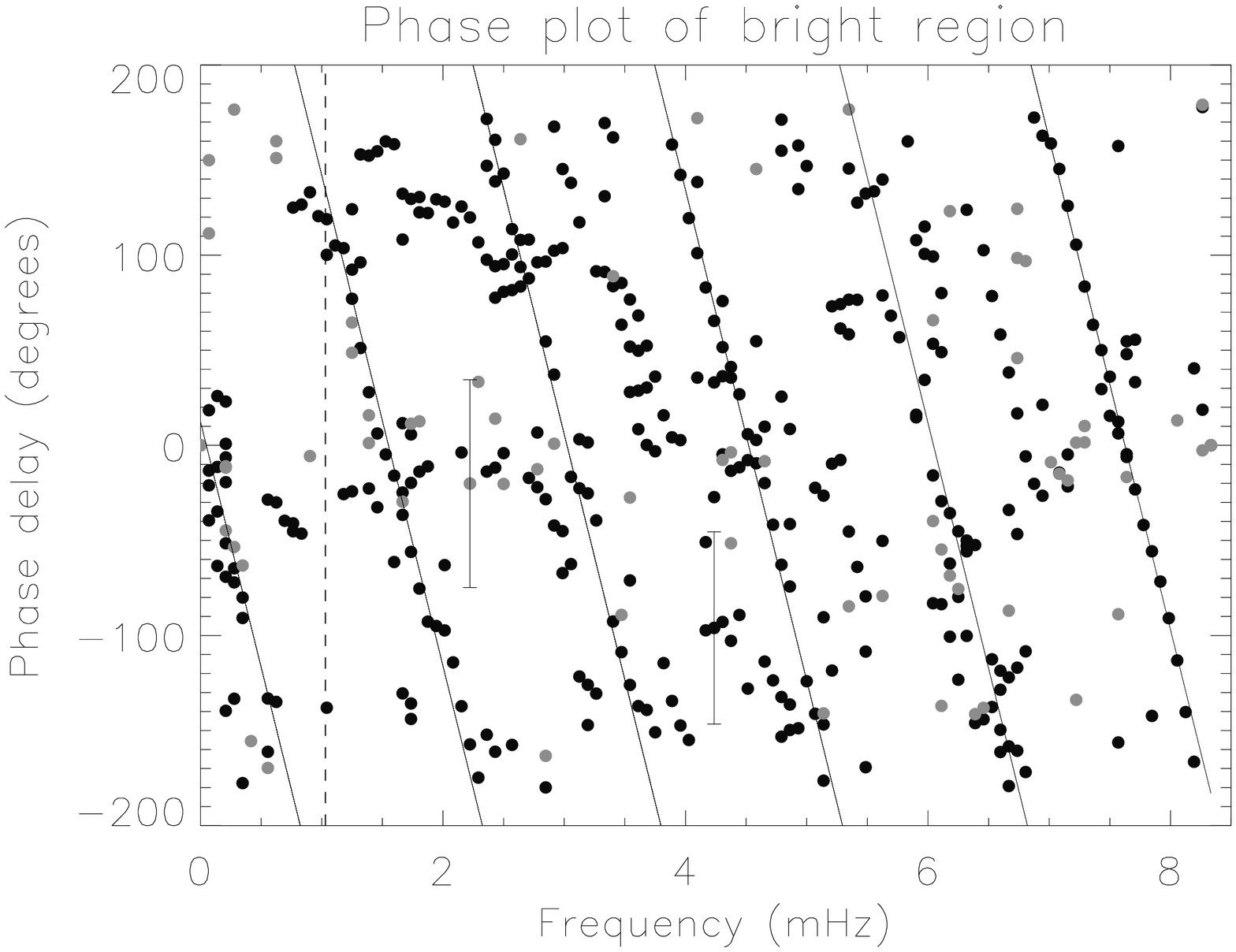}\hspace*{1cm}{\includegraphics[angle=90, width=8.7cm]{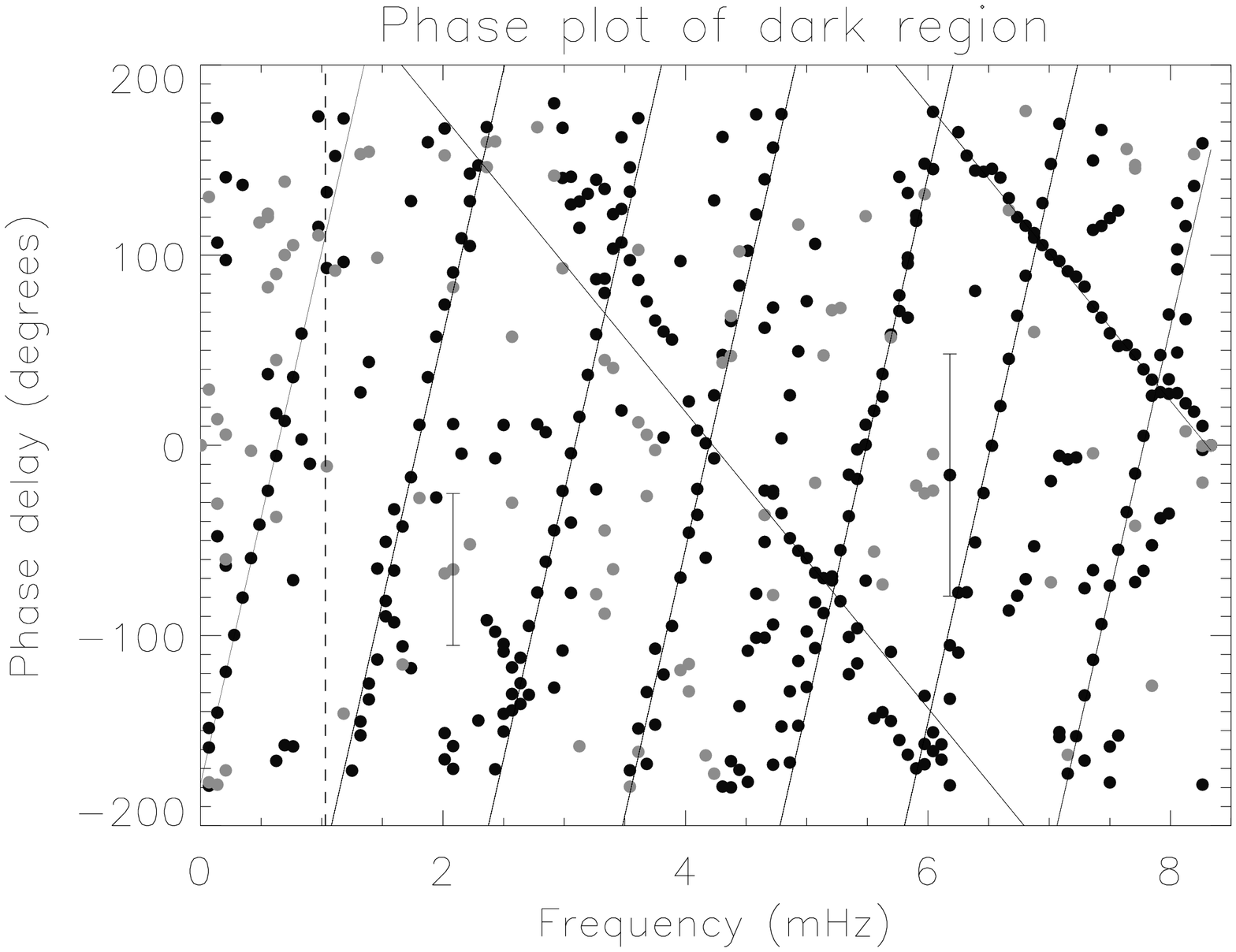}}
 \caption{Phase delays measured between the oscillations in the
 spectroscopic line pair for the bright (left panel) and dark (right panel) locations . The phases
 in radiant flux oscillations are shown in the grey circle symbols
 while that in LOS velocities are shown as the black circle
 symbols. Overplotted on each figure are black  parallel lines,
 corresponding to fixed time delays. The vertical dashed line drawn at 1.03 mHz indicates that some phase values below this could be
affected by solar rotation. Representative errors on the phase measurements are indicated by
 the error bars.}
 \label{fig:3a}
\end{figure*}
%-----------------------------------------------
\begin{figure*}[htbp]
\centering
%\hspace
 \includegraphics[angle=90, width=9.2cm]{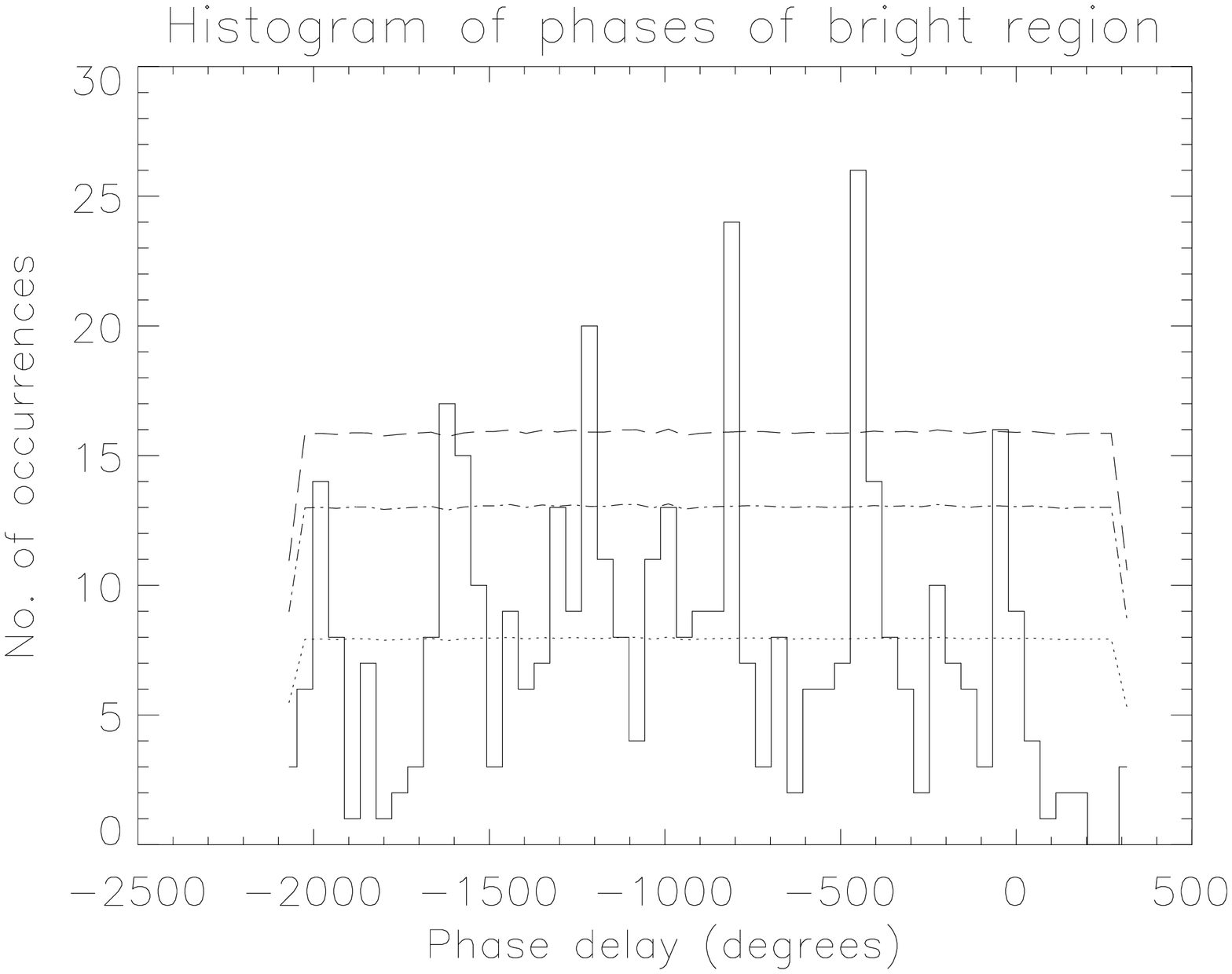}\includegraphics[angle=90, width=9.2cm]{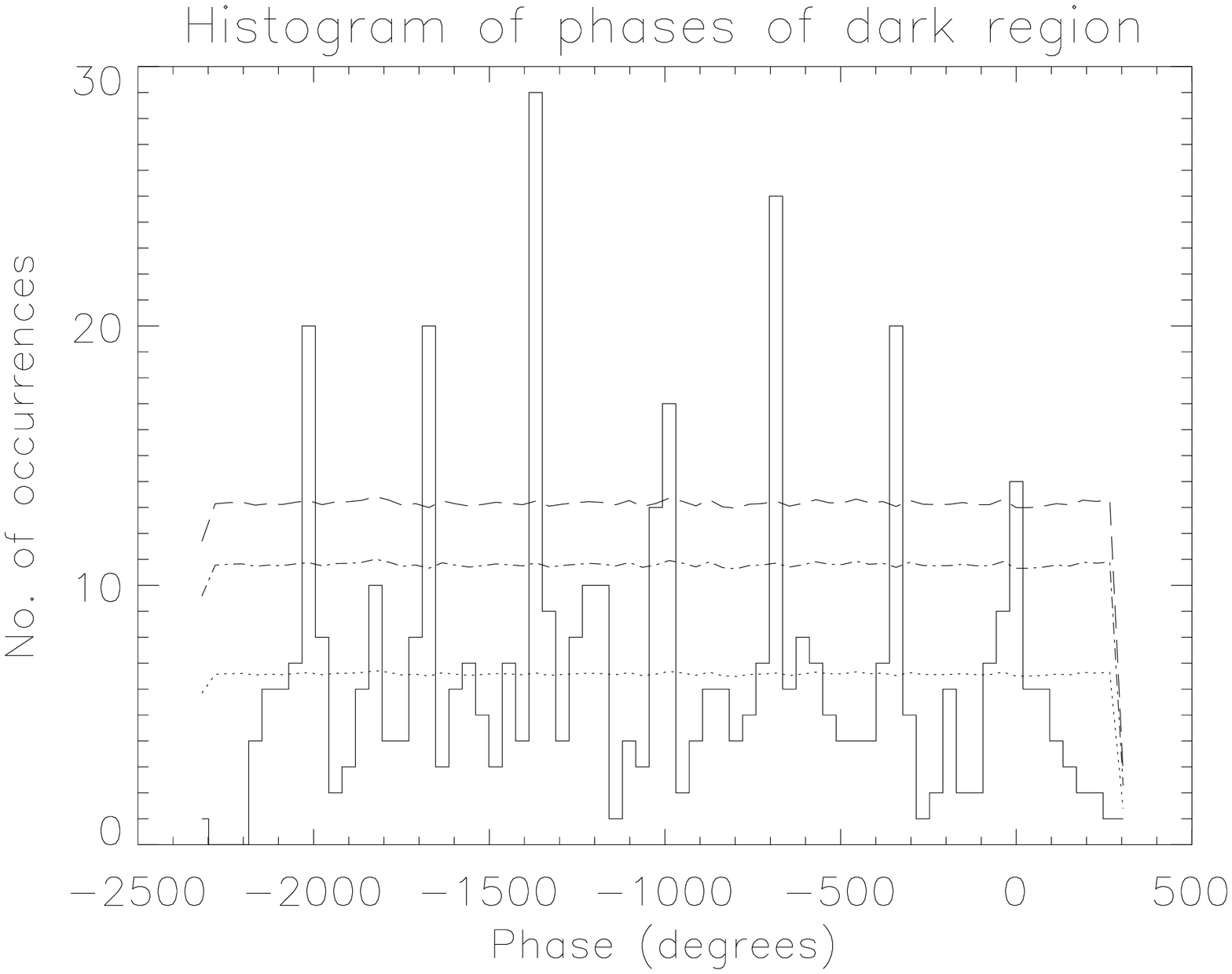}
\caption{Histogram showing the distribution of phase delay
  measurements as a function of frequency for the bright (left panel) and dark (right panel) locations for the downwardly propagating waves only. The horizontal
  dotted, dot-dashed and long dashed lines show the 68.3\%
  (i.e.,1$\sigma$), 90\% ($\approx$1.64$\sigma$) and
  95\% ($\approx$2$\sigma$) confidence levels, respectively,
  calculated using Monte-Carlo simulations with 5000 permutations.}
\label{fig:3aa}
\end{figure*}
%--------------------------------------------------------------------------
 Following the techniques outlined in \citet{O'Shea06}, we show in
Fig.~\ref{fig:3a}, the combined phase delay results for bright and
dark regions. 
The phase here is calculated for the line pair N~{\sc iv}
765--Ne~{\sc viii} $770$ in intensity and in relative velocity. If the
oscillations are due to magneto-acoustic waves, it is to be
expected that wave signatures will be present in both intensity
and velocity. For this reason, the calculated phases for
both of these quantities are plotted together, which also helps to improve the statistics. The phases from radiant flux oscillations are shown as the grey circle
symbols, while those from LOS velocities are shown as the black circle
symbols. We should point out here that we have approximately 50 network
and 50 internetwork pixels after binning. Each panel of Fig.5 has about
450 phases points, including intensity and velocity phase points. On
average, for each pixel position (along the slit) we have about 9 phase
points, that have a significance of 95.4\% and above, with more velocity
points as compared to intensity points. 
We can see immediately that the phases in Fig. ~\ref{fig:3a} are distributed between $-180$ and $180$. If there are fixed time delays
present between the oscillations in the different lines, as in
\citet{O'Shea06}, then we might expect the phases to line up along
inclined parallel lines as predicted by the phase equation (Eq.~1). We briefly summarise the method followed here.
In the phase plot (Fig. ~\ref{fig:3a}), we first identify the location where the strong clustering of phase points forming the straight line takes place e.g., a clear line of points between 7 and 8 mHz in the left panel of 
Fig.~\ref{fig:3a}. This line need not pass through the origin
i.e. point where the phase equals zero for $f$=0 as done in the
previous studies \citep{O'Shea06, O'Shea07}. We note, however, that
the first line does pass through this point, as expected. [For the plot
in the right-hand panel, showing mostly downward propagating waves,
the first line does not pass through the $f$=0 point. If we imagine,
however, a further line to the left of the first line, with the same
spacing in $f$, then this would likely pass through the zero point,
assuming that the errors in the phases are $\approx$50-60 degrees. 
After identifying the location, a straight line (between 7-8 mHz in
the left-hand plot) is fitted to these phase points. After fitting this line, all the phase points are shifted by a fixed amount which is determined from the slope of the fitted line, resulting in a transfrom which make the lines horizontal. Finally histograms are produced with these results (Fig.~\ref{fig:3aa}). Numerous peaks are present in the histogram. Now to check if these peaks are statistically significant or just due to a random distribution, we applied the Monte-Carlo technique, where the phase data were randomised 5000 times, to give estimates of the 'noise' (1 $\sigma$) level in the histogram. We have then calculated the 90 \% and 95.4 \% significance levels. So in the histogram, peaks above the 95.4 \% level are statistically more significant.
Furthermore, each of these peak indicates the presence of a straight line at that position. Hence, other parallel straight lines are drawn in the phase plot in Fig.~\ref{fig:3a} by using these positions.
 Thus, by fitting lines to the phase measurements in the phase plots, we estimate the slopes and therefore the
time delays between the line pair for the bright and dark regions. 
Looking at the left panel of Fig.~\ref{fig:3a} first, for the bright (`network')
locations, it is  clear that parallel lines are present in the
data. These lines of phase are more obvious in the measurements made
from the velocity oscillations. The implications of this result will be dealt with in the discussion section. 
The slope of these lines gives a time delay of $-717\pm 114$s. 
%To see if these lines are statistically significant or just due to a
%random distribution, we produce histograms from the phase distribution
%as in \citet{O'Shea06}. Examining Fig.~\ref{fig:3aa}, one can see that
%there are numerous histogram peaks above the 95\% significance
%level. These levels were produced using a Monte-Carlo technique,
%whereby the phase data was randomised 5000 times, to give estimates of
%the `noise' (1$\sigma$) level in the histograms. 
It is noticeable that
these histogram peaks are constantly spaced at $\approx$400$\degree$. As the
errors in the phases shown in  Fig.~\ref{fig:3a} are of the order of
$\approx$ $\pm$ 55 $\degree$, it is likely that what we see here is
really just the
expected variation of phases over 360$\degree$ for upwardly 
propagating waves. The consistent variation of
$\approx$400$\degree$ suggests, as in \citet{O'Shea06, O'Shea07},
that we may also be seeing a signature of cavity resonance. 
The result can indeed be interpreted as:
\begin{equation}
\Delta \phi=2\pi(f \pm n\Delta f)T
\label{eqn2}
\end{equation}
where $n$ is the order of the frequency shift, i.e., 0,1,2, etc., 
and $\Delta f$ is $f/9$ for the results found here. For example, for
the time delay of -717s (equivalent to a $f$ of 1.39 mHz), this shift,
$\Delta f$, of $f/9$ would give an exact phase spacing of
400$\degree$. 

%--------------------------------
In the dark locations (the `internetwork') 
(right panel of Fig.~\ref{fig:3a}), we find  two sets of parallel straight lines with opposite slope. The
presence of parallel lines with opposite slope indicates the presence
of both upward and downward propagating waves. This is the first time,
to our knowledge, that this has been reported in the literature. The
measured time delays in this location are $778\pm 133$s (downwardly
propagating), with the lines sloping towards the right, and $-216\pm
37.8$s (upwardly propagating), with the lines sloping towards the
left. In right panel of Fig.~\ref{fig:3aa}  we plot the
histograms corresponding to the downwardly  propagating
waves. Similar histograms have also been studied for upward
propagation but due to the lack of `lines' present in the phase no
conclusions could be drawn.
 Notice that in the right panel of Fig.~\ref{fig:3aa} the histogram
peaks are all above the 95\% significance level, indicating that it is
very unlikely that the lines of phase are due to noise. Notice
also that the spacing of the histogram peaks is not constant, but
varies between $\approx$300-340$\degree$. We suggest that this is the
expected 360$\degree$ phase difference for unimpeded downwardly
propagating waves, modified by the errors of $\approx$ $\pm$ 55 $\degree$ in
the phase measurements.

%%%%%%%%%%%%%%%%%%%%%%%%%%%%%%%%%%%%%%%%%%%%%%%%%%%

\begin{figure}
\centering
%\plotone{height.eps}
\includegraphics[angle=90, width=9cm]{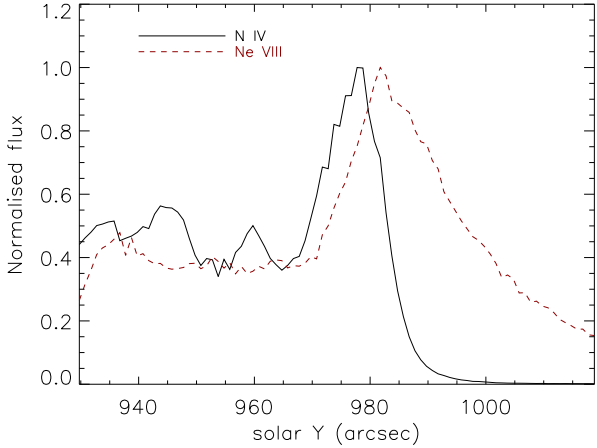}
\caption{Normalised flux distribution with height for the two lines. The solid line shows the
  distribution for N~{\sc iv}, while the dashed line shows the
  distribution for Ne~{\sc viii}. The location of maximum brightness
  for each line is used for determining the height difference.}
\label{fig:4}
\end{figure}

%%%%%%%%%%%%%%%%%%%%%%%%%%%%%%%%%%%%% Table one

\begin{table*}
\begin{small}
\caption{Propagation speeds measured in the polar coronal hole.}
 \begin{tabular}{ccccc}
  \hline\hline
 Height & \multicolumn{2}{c}{Delay time (s)} &
\multicolumn{2}{c}{Wave speed (km/s)} \\
 Diff. (Km) &  Bright location & Dark location & Bright location & Dark location \\
\hline
&  & $778.0\pm 133.5$ & & $3.7\pm 0.5$ \\
 \raisebox{1.5ex}{$2860$} & \raisebox{1.5ex}{$-717.0\pm 114.1$} & $-216.4\pm 37.8$ & \raisebox{1.5ex}{$-4.0\pm 0.6$} & $-13.2\pm 2.3$ \\
\hline     
%2 & $4095$ & $-278.4 \pm 32.07$ & $-2607 \pm 491.2$  &  $-14.71 \pm 1.695 $ & $-1.570 \pm 0.2959$ \\
%\hline
\label{tab:re_tab2}
 \end{tabular}
 \end{small}
%\caption{Propagation speeds measured in the analysed polar coronal holes.}

\end{table*}

%--------------------------------------------------

The measured time delays  may be used to
estimate propagation speeds for the waves assumed to be causing the
oscillations. In order to calculate the propagation speeds, one needs
information on the height difference in the atmosphere between the
different lines in the line pair. As described in \citet{O'Shea06}, 
we studied the variation of the summed flux over time
along the slit position near the limb (see Fig.~\ref{fig:4}). Noting the peak intensity location we have estimated the
 difference in their formation heights. So using this height
 difference and the measured time delays,
we show in Table~\ref{tab:re_tab2} the resulting calculated propagation speeds.
 The sound speed in the solar atmosphere can be expressed as function
of temperature alone \citep{Priest84} by the equation:
\begin{equation}
C_{s}=152 T^{1/2} m s^{-1}
\end{equation}
where $T$ is the temperature in K. 
From this, and using the formation
temperature of the lines,  we can estimate
the sound speeds to be 48 km s$^{-1}$ at the temperature of N~{\sc iv} (T $\approx$ $1 \times 10^{5}$)  and 120 km
s$^{-1}$ at the temperature of Ne~{\sc viii} (T $\approx$ $6.3 \times 10^{5}$). It can be clearly seen,
therefore, that the speeds we measured for different locations in the
coronal holes are subsonic.

\section{Conclusions}

Using the spectral diagnostic capabilities of the SUMER spectrograph,
and with a rigorous statistical approach, we have found evidence for 
propagating magneto-acoustic waves in polar coronal hole regions. For a typical bright enhanced network location
the  dominant period is $\approx$25 min. 
%Most earlier studies of wave detection in coronal holes have been
%done either using imaging instruments or in very specific locations
%only. These  earlier results indicate presence of power in specific
%frequency only. 
Phase measurements were carried out over the full range of frequencies
measurable in our data, not just at the dominant period. From these
phase measurements, we find evidence that the waves producing the
oscillation signatures propagate both upwards and downwards.
This is the first time, to our knowledge, that this has been reported
in the literature. The dark
(`internetwork') areas in pCH show evidence of both upward and
downward 
propagating waves, while bright (`network') areas, by comparison, show
only upwardly propagating waves. This may indicate different physical
processes occurring in the `internetwork' and `network' regions. From
a study of the height variations of the sizes of chromospheric and 
transition-region features in a small coronal hole and the adjacent
quiet Sun, \cite{tian2008} concluded that loops are much lower in the
coronal hole than in the quiet Sun region. From a study of the height 
variation of the Half Width at Half Maxima of $\left\vert
B_{z}/B\right\vert$, they further suggested that an open magnetic
structure expands through the upper transition region and lower corona
more strongly in coronal holes than in the quiet Sun. We may conjecture
that the internetwork regions within coronal holes are composed of
low-lying coronal loops, where waves can travel both upwardly and
downwardly, whereas the network regions are filled with more open,
funnel like structures where only upward propagation is possible. 
Recently, \citet{Gomory06} reported detection of compressive waves that propagate downward from transition region to the chromosphere in a particular chromospheric network in a quite Sun region. They propose a coronal source
 of these compressive waves resulting from a nano-flare scenario. In our case clearly we see presence of downward propagation only for internetwork location. One can conjecture that the magnetic topology in quiet sun could be slightly different than in the coronal holes, with presence of more low-lying loop systems in the internetwork regions of coronal loops. 

Plotting the phases as histograms, we saw a
possible cavity resonance effect on the waves, e.g., $\Delta f$s of $f/9$, inferred in reference to Eqn. ~\ref{eqn2}, as well as the expected
unimpeded propagation of the waves, e.g., a measured phase difference
of 360$\degree$ between the different parallel lines of phases. We
note that the speeds  we derived (cf. Table~\ref{tab:re_tab2}) are
lower than those found by \citet{O'Shea07} in polar coronal holes
on-disc, and much lower than those found by \citet{O'Shea06}
off-limb. However, the wave speeds measured by \citet{O'Shea07} were
for lines with high formation temperature and, hence, higher up
in the atmosphere as compared to our lines. 
 We hypothesise that at lower heights the waves have lower speeds, but 
the speeds will increase with height in the solar atmosphere.
 We found the propagation speeds to be
consistent with slow magneto-acoustic waves. We should also point out
that statistically we find more velocity oscillations than intensity
oscillations in this study. For compressional types of waves, we
should have signatures in both. So if there are only velocity
oscillations present in some locations this implies the presence of transverse-type waves, like Alfv\'{e}nic. Recent studies from HINODE by 
\cite{dePontie2007} suggest that the chromosphere is permeated by
Alfv\'{e}n waves with strong amplitudes of the order of 10 to 25 km s$^{-1}$ and periods of 100 to 500s. \citet{Tsuneta08} have further conjectured that the vertical flux tubes originating from the kG patches in coronal holes will fan out in the lower atmosphere of the coronal holes and will serve as efficient chimney for the propagation of Alfv\'{e}n waves. These waves may carry enough energy flux for the acceleration of the solar wind. So, it is quite possible that some of these Alfv\'{e}nic waves have also been detected in our statistical analysis as we see more velocity points as compared to intensity points in the phase plots. 
\begin{acknowledgements}                 

Research at Armagh Observatory is grant-aided
by the N.~Ireland Dept. of Culture, Arts and Leisure.  This work 
was supported by a Royal Society/British Council and  STFC grant PP/D001129/1.
The SUMER project is financially supported by DLR, CNES, NASA, and the ESA PRODEX programme (Swiss contribution). We thank the Referee for a careful reading of the manuscript and for very valuable suggestions, which has improved the quality of the article.
\end{acknowledgements}
%%%%%%%%%%%%%%%%%%%%%%%%%%%%%%%%%%%%%%%%%%%% References %%%%%%%%%%%%%%%%%%%%%%%%%%%%%%
\bibliographystyle{aa}

\begin{thebibliography}{}
\bibitem[Athay \& White (1979)]{Athay79}
Athay, R. G., \&  White, O. R.
1979, \apj , 229, 1147

\bibitem[Banerjee et al. (2001)]{Banerjee01}
Banerjee, D., O'Shea, E., Doyle, J. G., \& Goossens, M.
2001, \aap, 380, L39

\bibitem[Dammasch et al. (1999)]{Dammasch99}
Dammasch, I. E., Wilhelm, K., Curdt, W., Hassler, D. M.
1999, \aap, 346, 285

\bibitem[De Pontieu et al. (2007)]{dePontie2007}
De Pontieu, B., McIntosh, S. W., Carlsson, M., Hansteen, V. H., Tarbell, T. D., Schrijver, C. J., Title, A. M., Shine, R. A., Tsuneta, S., Katsukawa, Y., Ichimoto, K., Suematsu, Y., Shimizu, T., Nagata, S. 2007, Science,  318, 1574

\bibitem[DeForest \& Gurman (1998)]{DeForest98}
DeForest, C. E., \& Gurman, J. B.
1998, \apjl, 501, L217

%\bibitem[Delaboudiere et al. (1995)]{Delab95}
%Delaboudiere, J.-P., Artzner, G.E., Brunaud, J., \etal, 1995, \solphys, 162, 291

\bibitem[Doyle et al. (1998)]{Doyle98}
Doyle, J. G., Van Den Oord, G. H. J., O'Shea, E., Banerjee, D.
1998, \solphys, 181, 51

\bibitem[Doyle et al. (1999)]{Doyle99}
Doyle, J. G., Van Den Oord, G. H. J., O'Shea, E., Banerjee, D.
1999,  \aap, 347, 335
\bibitem[Gomory et al. (2006)]{Gomory06}
Gomory, P., Rybak, J., Kucera, A., Curdt, W., Wohl, H.
2006, \aap, 448, 1169

\bibitem[Jenkins \& Watts (1968)]{Jenkins68} 
Jenkins, G.M., \& Watts, D.G.
1968, Spectral analysis and its applications, Holden-Day, oakland

\bibitem[Krieger et al. (1973)]{Krieger73}
Krieger, A., Timothy, A. F., \& Roelof, E. C.
1973, \solphys, 29, 505

\bibitem[Munro \& Withbroe (1972)]{Munro72}
Munro, R. H., \& Withbroe, G. L.
1972, \apj, 176, 511

\bibitem[Ofman et al.(1997)]{Ofman97}
Ofman, L., Romoli, M., Poletto, G., Noci, G., \& Kohl, J. L.
1997, \apjl, 491, L111

\bibitem[Ofman et al.(2000)]{Ofman00}
Ofman, L., Romoli, M., Poletto, G., Noci, G., \& Kohl, J. L.
2000, \apj, 529, 592

\bibitem[O'Shea et al.(2001)]{O'Shea01}
O'Shea, E., Banerjee, D., Doyle, J. G., Fleck, B., \& Murtagh, F.
2001, \aap, 368, 1095

\bibitem[O'Shea, Banerjee \& Doyle (2006)]{O'Shea06}
O'Shea, E., Banerjee, D., \& Doyle, J. G.
2006, \aap, 452, 1059

\bibitem[O'Shea, Banerjee \& Doyle (2007)]{O'Shea07}
O'Shea, E., Banerjee, D., \& Doyle, J. G.
2007, \aap, 463, 713

\bibitem[Popescu, Doyle \& Xia (2004)]{Popescu04}
Popescu, M. D., Doyle, J. G., \& Xia, L. D.
2004, \aap, 421, 339 

\bibitem[Popescu et al. (2005)]{Popescu05}
Popescu, M. D., Banerjee, D., O'Shea, E., Doyle, J. G., \& Xia, L. D.
2005, \aap, 442, 1087

\bibitem[Priest (1984)]{Priest84}
Priest, E. R.
1984, Solar Magnetohydrodynamics, Dordrecht: Reidel

\bibitem[Reeves (1976)]{Reeves1976}
Reeves, E. M. 1976, \solphys, 46, 53

\bibitem[Tian et al. (2008)]{tian2008}
Tian, H., Marsch, E., Tu, C.-Y., Xia, L.-D., He, J.-S. 2008 \aap, 482, 267

\bibitem[Timothy, Krieger \& Vaiana (1975)]{Timothy75}
Timothy, A. F., Krieger, A. S., \& Vaiana, G. S.
1975, \solphys, 42, 135

\bibitem[Tsuneta et al. (2008)]{Tsuneta08}
Tsuneta, S. et al.
arXiv:0807.4631v1

\bibitem[Torrence \& Compo (1998)]{Torrence98}
Torrence, C., \& Compo, G. P.
1998, Bull. Amer. Meteor. Soc., 79, 61

\bibitem[Tu et al.(2005)]{Tu05}
Tu, C.-Yi, Zhou, C., Marsch, E., Xia, L., Zhao, L., Wang, J.-X, \& Wilhelm, K.
2005, Science, 308, 519

\bibitem[Wilhelm et al. (1995)]{Wilhelm95}
Wilhelm, K., Curdt, W., Marsch, E., \etal, 1995, \solphys, 162, 189

\bibitem[Wilhelm et al. (2000)]{Wilhelm00}
Wilhelm, K., Dammasch, I. E., Marsch, E., \& Hassler, D. M.
2000, \aap, 353, 749

\bibitem[Xia, Marsch \& Curdt (2003)]{Xia03}
Xia, L. D., Marsch, E., \& Curdt, W.
2003, \aap, 399, L5

\bibitem[Xia, Marsch \& Wilhelm (2004)]{Xia04}
Xia, L. D., Marsch, E., \& Wilhelm, K.
2004, \aap, 424, 1025

%\bibitem[Xia et al. (2005)]{Xia05}
%Xia, L. D., Popescu, M. D., Doyle, J. G., \& Giannikakis, J.
%2005, \aap, 438, 1115

\end{thebibliography}

\end{document}